%% file: etapt_pp13_CDS.tex
\documentclass[ALICE,manyauthors]{cernphprep}

\usepackage[comma,square,numbers,sort&compress]{natbib}
\usepackage{hyperref}
\usepackage{lineno}
\usepackage{multirow}
\usepackage{color}

\input{commands}

\begin{document}%

\begin{titlepage}
\PHyear{2015}
\PHnumber{270}      
\PHdate{29 September}  
%

\title{Pseudorapidity and transverse-momentum distributions of\\charged particles in proton-proton collisions at $\mathbf{\sqrt{\textit s}}$ = 13 TeV}
\ShortTitle{$\eta$ and \pt distributions of charged particles in pp at \sqrts = 13 TeV}   

\Collaboration{ALICE Collaboration\thanks{See Appendix~\ref{app:collab} for the list of collaboration members}}
\ShortAuthor{ALICE Collaboration} 

\begin{abstract}
\input{abstract}
\end{abstract}
\end{titlepage}
\setcounter{page}{2}

\input{paper}               
%
%

\newenvironment{acknowledgement}{\relax}{\relax}
\begin{acknowledgement}
\section*{Acknowledgements}
\input{acknowledgements.tex}    
\end{acknowledgement}

\bibliographystyle{utphys}   
\bibliography{bibliography}

\newpage
\appendix
\section{The ALICE Collaboration}
\label{app:collab}
\input{Alice_Authorlist_2015-Sep-24.tex}  
\end{document}

%% file: commands.tex
\newcommand{\pt}{\ensuremath{p_{\rm T}}\xspace}
\newcommand{\sqrts}{\ensuremath{\sqrt{s}}\xspace}
\newcommand{\dndeta}{\ensuremath{{\rm d}N_{\rm ch}/{\rm d}\eta}\xspace}
\newcommand{\dndpt}{\ensuremath{{\rm d}N_{\rm ch}/{\rm d}\pt}\xspace}
\newcommand{\etaless}[1]{\ensuremath{\left|\eta\right| < #1}\xspace}

\newcommand{\MeVc}{MeV/$c$\xspace}
\newcommand{\GeVc}{GeV/$c$\xspace}
\newcommand{\pp}{\ensuremath{\rm pp}\xspace}
\newcommand{\pPb}{p--Pb\xspace}
\newcommand{\PbPb}{Pb--Pb\xspace}

\newcommand{\inelgtzero}{INEL$>$0\xspace}

\newcommand{\SPS}{\ensuremath{\rm Sp\bar{p}S}\xspace}

\newcommand{\average}[1]{\ensuremath{\langle #1 \rangle}\xspace}
\newcommand{\mpt}        {\ensuremath{\langle\pt\rangle}\xspace}
\newcommand{\nch}        {\ensuremath{N_\mathrm{ch}}\xspace}
\newcommand{\mnch}      {\ensuremath{\langle\nch\rangle}\xspace}
\newcommand{\nchacc}        {\ensuremath{N_\mathrm{ch}^\mathrm{acc}}\xspace}
\newcommand{\mnchacc}      {\ensuremath{\langle\nchacc\rangle}\xspace}

\newcommand{\DNDETAINEL}{5.31~$\pm$~0.18\xspace}
\newcommand{\DNDETAINELGTZERO}{6.46~$\pm$~0.19\xspace}
\newcommand{\DNDETAINELGTZEROONE}{6.61~$\pm$~0.20\xspace}

%% file: abstract.tex
The pseudorapidity ($\eta$) and transverse-momentum (\pt) distributions of charged particles produced in proton-proton collisions are measured at the centre-of-mass energy \sqrts~=~13~TeV.
The pseudorapidity distribution in \etaless{1.8} is reported for inelastic events and for events with at least one charged particle in \etaless{1}.
The pseudorapidity density of charged particles produced in the pseudorapidity region \etaless{0.5} is \DNDETAINEL and \DNDETAINELGTZERO for the two event classes, respectively.
The transverse-momentum distribution of charged particles is measured in the range 0.15~$<$~\pt~$<$~20~\GeVc and \etaless{0.8} for events with at least one charged particle in \etaless{1}.
The evolution of the transverse momentum spectra of charged particles is also investigated as a function of event multiplicity.
The results are compared with calculations from PYTHIA and EPOS Monte Carlo generators.

%% file: paper.tex
\input{introduction.tex}
\input{detector.tex}
\input{analysis.tex}
\input{systematics.tex}
\input{results.tex}
\input{conclusions.tex}

%% file: introduction.tex
\section{Introduction}\label{sec:introduction}

After a two-year long shutdown, the Large Hadron Collider (LHC) at CERN restarted its physics programme in June 2015 with proton-proton collisions at \sqrts~=~13~TeV, the highest centre-of-mass energy reached so far in laboratory.
The measurement of the inclusive production of charged hadrons in high-energy proton-proton interactions is a key observable to characterise the global properties of the collision, in particular whenever the collision energy increases significantly. 
Particle production at collider energies originates from the interplay of perturbative (hard) and non-perturbative (soft) QCD processes. Soft scattering processes and parton hadronisation dominate the bulk of particle production at low transverse momenta and can only be modelled phenomenologically. Hence, these measurements provide constraints for a better tuning of models and event generators for hadron-collider and cosmic-ray physics~\cite{d'Enterria:2011kw}. 

We present the pseudorapidity ($\eta$) and transverse-momentum (\pt) distributions of primary charged particles measured in proton-proton collisions at the centre-of-mass energy \sqrts~=~13~TeV with the ALICE detector~\cite{Aamodt:2008zz} at the LHC~\cite{Evans:2008zzb}. Primary particles are defined as prompt particles produced in the collisions, including all decay products, with the exception of those from weak decays of strange particles. Similar measurements have been performed by ALICE in proton-proton (pp), proton-lead (p--Pb) and lead-lead (Pb--Pb) collisions collected during the previous LHC run at lower energies~\cite{Aamodt:2010ft,Aamodt:2010pp,LongMultiPaper,ALICE:2012xs,Aamodt:2010pb,Aamodt:2010my,Abelev:2013ala,ALICE:2012mj,Abelev:2014dsa,Aamodt:2010jd,Abelev:2012eq}.
The pseudorapidity distribution is measured at central rapidity in \etaless{1.8}.  
The measurements reported here have been obtained for inelastic events (INEL) and events having at least one charged particle produced with \pt~$>$~0 in the pseudorapidity interval \etaless{1} (\inelgtzero).
Similar results were recently published by the CMS Collaboration for INEL events~\cite{Khachatryan:2015jna}.
The transverse-momentum distribution of charged particles is measured in the range 0.15~$<$~\pt~$<$~20~\GeVc and \etaless{0.8} for \inelgtzero events.
The evolution of the transverse momentum spectra of charged particles is also investigated as a function of event multiplicity.
The data have been compared to calculations from models commonly used at the LHC. 

%% file: detector.tex
\hyphenation{offline}

\section{The ALICE detector and data collection}\label{sec:detector}

A comprehensive description of the ALICE experimental setup can be found in~\cite{Aamodt:2008zz,Abelev:2014ffa}. The main detectors utilised for the analysis presented here are the Inner Tracking System (ITS), the Time-Projection Chamber (TPC), the V0 counters and the ALICE Diffractive (AD) detector.
The ITS and TPC detectors, which are located inside a solenoidal magnet providing a magnetic field of 0.5~T, are used for primary-vertex and track reconstruction. The V0 counters and the AD detector are employed for triggering and for background suppression.

The ITS is composed of six cylindrical layers of high-resolution silicon tracking detectors. The innermost layers consist of two arrays of hybrid Silicon Pixel Detectors (SPD) located at radii 3.9 and 7.6~cm from the beam axis and covering respectively \etaless{2.0} and \etaless{1.4} for particles emerging from the nominal interaction point ($z$~=~0~cm).
The TPC is a large cylindrical drift detector of radial and longitudinal size of about $85 < r < 250$ cm and $-250 < z < 250$ cm, respectively. The active volume of nearly 90~m$^3$ is filled with an Ar-CO$_2$ (88-12\%) gas mixture and is divided in two halves by a central high-voltage membrane maintained at $-100$ kV. The two end-caps are each equipped with 36 multi-wire proportional chambers with cathode pad readout, comprising a total of 558000 readout channels.
The V0 counters are two scintillator hodoscopes placed on either side of the interaction region at $z = 3.3$~m and $z = -0.9$~m, covering the pseudorapidity regions $2.8 < \eta < 5.1$ and $-3.7 < \eta < -1.7$, respectively.
The AD detector was integrated in ALICE during the LHC shutdown before Run 2 to enhance the capabilities of the experiment to tag diffractive processes and low \pt events~\cite{forwardp}. It consists of two double layers of scintillation counters placed far from the interaction region, on both sides: one in the ALICE cavern at $z = 17.0$~m and one in the LHC tunnel at $z = -19.5$~m. The pseudorapidity coverage of the two AD arrays is $4.8 < \eta < 6.3$ and $-7.0 < \eta < -4.9$, respectively.

The data were collected after the startup of LHC Run 2 in June 2015. Beams consisting of 39 bunches were circulating in the machine, with about $8 \times 10^{9}$ protons per bunch. In the ALICE interaction region, 15 pairs of bunches were colliding, leading to a luminosity of about $5 \times 10^{27} \rm cm^{-2} s^{-1}$. This value corresponds to a rate of about $350$~Hz for inelastic proton-proton collisions. The probability that a recorded event contains more than one collision was estimated to be around $10^{-3}$, which is consistent with the fraction of events containing more than one distinct vertex and tagged as pileup. The luminous region had an RMS width of about 5~cm in the $z$ direction and about 85~$\mu$m in the transverse direction.
The data were collected using a minimum-bias trigger requiring a hit in either the V0 scintillators or in the AD arrays.
The events were recorded in coincidence with signals from two beam pick-up counters each positioned on either side of the interaction region to tag the arrival of proton bunches from both directions. Control triggers taken for various combinations of beam and empty buckets were used to measure beam-induced and accidental backgrounds. The contamination from background events is removed {{offline}} by using the timing information from the V0 and the AD detectors, which have a time resolution better than 1 ns. Background events are also rejected by exploiting the correlation between the number of clusters of pixel hits and the number of tracklets (short track segments pointing to the primary vertex) in the SPD. From the analysis of control triggers it is estimated that the remaining background fraction in the sample is less than $10^{-4}$ and can be neglected.

%% file: analysis.tex
\section{Event selection and data analysis}\label{sec:analysis}

About 1.5 million events pass the minimum-bias selection criteria.
Events used for the data analysis are further required to have a valid reconstructed vertex within $\left|z\right| <$~10~cm. All corrections are calculated using a sample of about 4 million Monte Carlo events from the PYTHIA~6~\cite{Sjostrand:2006za} (Perugia-2011~\cite{Skands:2010ak}) event generator with particle transport performed via a GEANT3~\cite{Brun:1994aa} simulation of the ALICE detector.

The analysis technique employed for the measurement of the charged-particle pseudorapidity distribution is based on the reconstruction of tracklets, which are built using the position of the reconstructed primary vertex and two hits, one on each SPD layer. Details on the algorithm for tracklet reconstruction are described in~\cite{Aamodt:2010ft}. This technique effectively allows to reconstruct charged particles with \pt above the 50~\MeVc cut-off determined by particle absorption in the material.
The charged-particle pseudorapidity density is obtained from the measured distribution of tracklets ${\rm d}N_{\rm tracklets}/{\rm d}\eta$ as ${\rm d}N_{\rm ch}/{\rm d}\eta = \alpha (1 - \beta) {\rm d}N_{\rm tracklets}/{\rm d}\eta$. The correction $\alpha$ accounts for the acceptance and efficiency for a primary particle to produce a tracklet, while $\beta$ is the contamination of reconstructed tracklets from combinations of hits not produced by the same primary particle. Both correction factors are determined as a function of the $z$ position of the primary vertex and the pseudorapidity of the tracklet from detector simulations and are found to be on average 1.5 and 0.01, respectively. The vertex position requirement results in an effective \etaless{1.8} coverage.
Differences in strange-particle content between data and simulations, observed at lower beam energies~\cite{Adam:2015qaa,Abelev:2012jp}, are taken into account by scaling the strangeness production in the Monte Carlo event sample by a factor 1.85 (strangeness correction), resulting in a further contamination correction of about 1\%.

The transverse-momentum distribution is measured from tracks reconstructed using the information from the ITS and TPC detectors. Candidate tracks are selected with cuts on the number of space points used for tracking and on the quality of the track fit, as well as on the distance of closest approach to the reconstructed vertex. Details on the track-reconstruction algorithm and quality cuts can be found in~\cite{Abelev:2012eq,Abelev:2013ala,ALICE:2012mj}. The requirements applied for track selection result in an effective \etaless{0.8} acceptance.
The efficiency for track reconstruction and selection depends on the particle type and it is known that PYTHIA~6 does not reproduce correctly the particle fractions measured at \sqrts = 7 TeV.
A reweighting of the Monte Carlo efficiencies for each species with the relative abundances measured in minimum-bias pp collisions at \sqrts = 7 TeV~\cite{Adam:2015qaa,Abelev:2012jp} is performed.
The overall primary charged-particle reconstruction efficiency for \etaless{0.8} increases sharply from 34\% at 150 \MeVc, reaches 73\% at 0.8 \GeVc, decreases moderately to 67\% for \pt = 2 \GeVc and rises again to reach a saturation value of 74\% at 10 \GeVc.
The minimum around 2 \GeVc arises due to the azimuthal segmentation of the TPC readout chambers. Tracks of moderate \pt, which may not have enough hits in adjacent azimuthal sectors, do not pass the selection criteria.
Finally, the residual contamination from secondary particles is subtracted from the spectrum; this contamination, estimated from Monte Carlo simulations, is 7\% for our lowest \pt bin and decreases below 1\% for \pt~$>$~2 \GeVc.

%% file: systematics.tex
\section{Systematic uncertainties}\label{sec:systematics}

\begin{table}[t]
  \centering
  \renewcommand{\arraystretch}{1.2}
  \renewcommand{\tabcolsep}{15pt}
  \begin{tabular}{l|lr|lr}
    & \multicolumn{2}{c|}{\dndeta} & \multicolumn{2}{c}{\dndpt} \\
    & INEL & \inelgtzero & 0.15 & 20 \GeVc\\
    \hline
    Background events and pileup & \multicolumn{2}{c|}{negligible} & \multicolumn{2}{c}{negligible} \\
    Normalisation & 2.8 & 2.3 & \multicolumn{2}{c}{2.3} \\
    \hline
    Detector acceptance and efficiency & \multicolumn{2}{c|}{1.5} & 1.8 & 5.6 \\
    Material budget & \multicolumn{2}{c|}{0.1} & 1.5 & 0.2 \\
    Track(let) selection criteria & \multicolumn{2}{c|}{negligible} & 1.5 & 3.0 \\
    Particle composition & \multicolumn{2}{c|}{0.2} & 0.3 & 2.4 \\
    Weak decays of strange hadrons & \multicolumn{2}{c|}{0.5} & 3.4 & 0.4 \\
    Zero-\pt extrapolation & \multicolumn{2}{c|}{1.0} & \multicolumn{2}{c}{not applicable} \\
    \hline
    Total ($\eta$, \pt dependent) &  \multicolumn{2}{c|}{1.9} & 4.4 & 6.8 \\
    \hline
    Total & 3.4 & 3.0 & 5.0 & 7.2 \\
  \end{tabular}
  \caption{Summary of the relative systematic uncertainties (expressed in \%) contributing to the measurement of the charged-particle pseudorapidity and transverse-momentum distributions. The values for the \dndeta analysis are reported separately for the INEL and \inelgtzero classes. For the \dndpt analysis the \pt dependence is summarised with the values at 0.15 and 20 \GeVc for the \inelgtzero class.}
  \label{tab:systematics}
\end{table}

A summary of the contributions to the relative systematic uncertainties of the charged-particle pseudorapidity and transverse-momentum distributions is reported in Tab.~\ref{tab:systematics}.

One of the main contributions to the normalisation of the results comes from the limited knowledge of cross-sections and kinematics of diffractive processes. For proton-proton collisions at \sqrts = 13 TeV there is not yet any experimental information available about diffractive processes, therefore trigger and event-selection efficiency corrections are solely based on previous experimental data at lower collision energies and simulations with Monte Carlo event generators.
The corresponding systematic uncertainty has been evaluated by varying the fractions of single-diffractive (SD) and double-diffractive (DD) events produced by PYTHIA~6 (Perugia-2011) by $\pm$50\% of their nominal values at \sqrts = 13 TeV. The resulting contribution to the systematic uncertainties for INEL and \inelgtzero events is estimated to be about 2\% and 1.2\%, respectively.
To estimate systematic uncertainties associated to the model dependence of the normalisation correction we employed PYTHIA~8~\cite{Sjostrand:2007gs} (Monash-2013~\cite{Skands:2014pea}), which shows large differences both in the multiplicity and transverse-momentum distributions of charged particles with respect to PYTHIA~6, especially in diffractive events~\cite{Navin:2010kk}.
A difference of about 0.4\% and 2\% is observed for INEL and \inelgtzero events, respectively.
Finally, an uncertainty of 2\% has been estimated by varying the offline event-selection criteria applied to the trigger detectors which only affects the normalisation of the INEL sample.

The systematic uncertainties for the transverse-momentum distribution analysis are evaluated in a similar way as in previous analyses of \pp \cite{Aamodt:2010my,Abelev:2013ala}, \pPb \cite{ALICE:2012mj,Abelev:2014dsa}, and \PbPb \cite{Abelev:2012eq} data.
The dominant sources of uncertainty are the track selections, the efficiency corrections and, for low \pt, the contamination from weak decays of strange hadrons.
The systematic uncertainties for the pseudorapidity distribution analysis are discussed in the following.
The uncertainty in detector acceptance and efficiency is estimated to be about 1.5\%, determined from the change of the multiplicity at a given $\eta$ by varying the range of the $z$ position of the vertex and performing the measurement in different runs. 
The material budget in the ALICE central barrel \etaless{1} is known with a precision of about 5\%~\cite{Abelev:2014ffa}. The corresponding systematic uncertainty, obtained by varying the material budget in the simulation, is estimated to be about 0.1\% and is negligibly small compared to the other sources.
The sensitivity to tracklet selection criteria was estimated varying the selection requirements and is negligible.
The uncertainty due to the particle composition is estimated to be about 0.2\% and was determined by changing the relative fractions of charged kaons and protons with respect to charged pions produced by the Monte Carlo generator by $\pm$30\%.
The uncertainty resulting from the subtraction of the contamination from weak decays of strange hadrons is estimated to amount to about 0.5\% by varying the strangeness correction by $\pm$30\%.
The uncertainty due to the correction down to zero \pt is estimated to be about 1\% by varying the amount of particles below the 50 \MeVc low-\pt cutoff by $^{+100}_{-50}$\%.

%% file: results.tex
\section{Results}\label{sec:results}

\begin{figure}[t]
  \centering
  \includegraphics[width=0.6\linewidth]{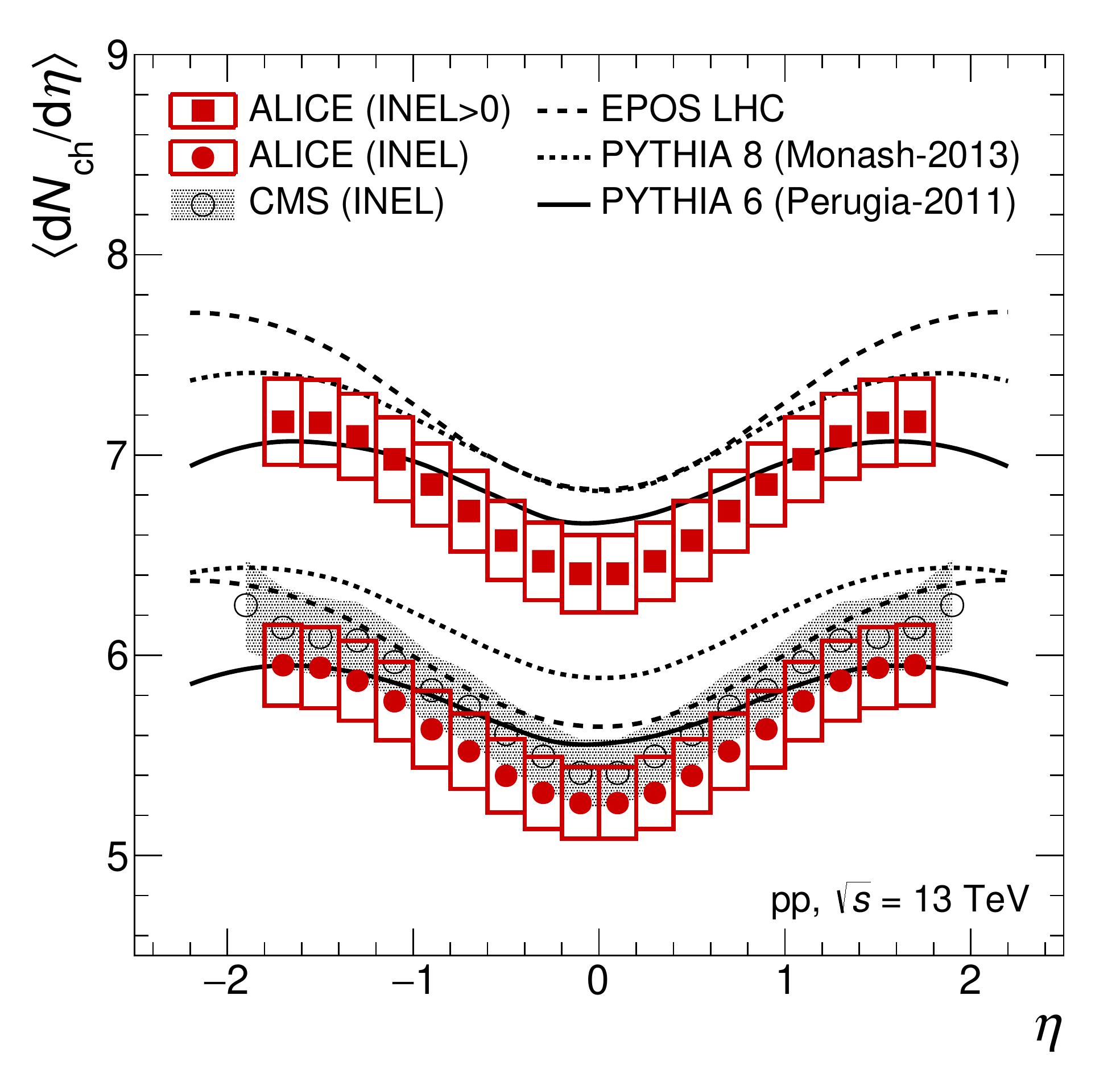}
\caption{Average pseudorapidity density of charged particles as a function of $\eta$ produced in pp collisions at \sqrts = 13 TeV. The ALICE results are shown in the normalisation classes INEL and \inelgtzero and compared to Monte Carlo calculations~\cite{Sjostrand:2006za,Skands:2010ak,Sjostrand:2014zea,Skands:2014pea,Drescher:2000ha,Pierog:2013ria} and to the results from the CMS Collaboration~\cite{Khachatryan:2015jna}. The uncertainties are the quadratic sum of statistical and systematic contributions.}
\label{fig:dndeta}
\end{figure}

\begin{figure}[t]
  \centering
  \includegraphics[width=0.6\linewidth]{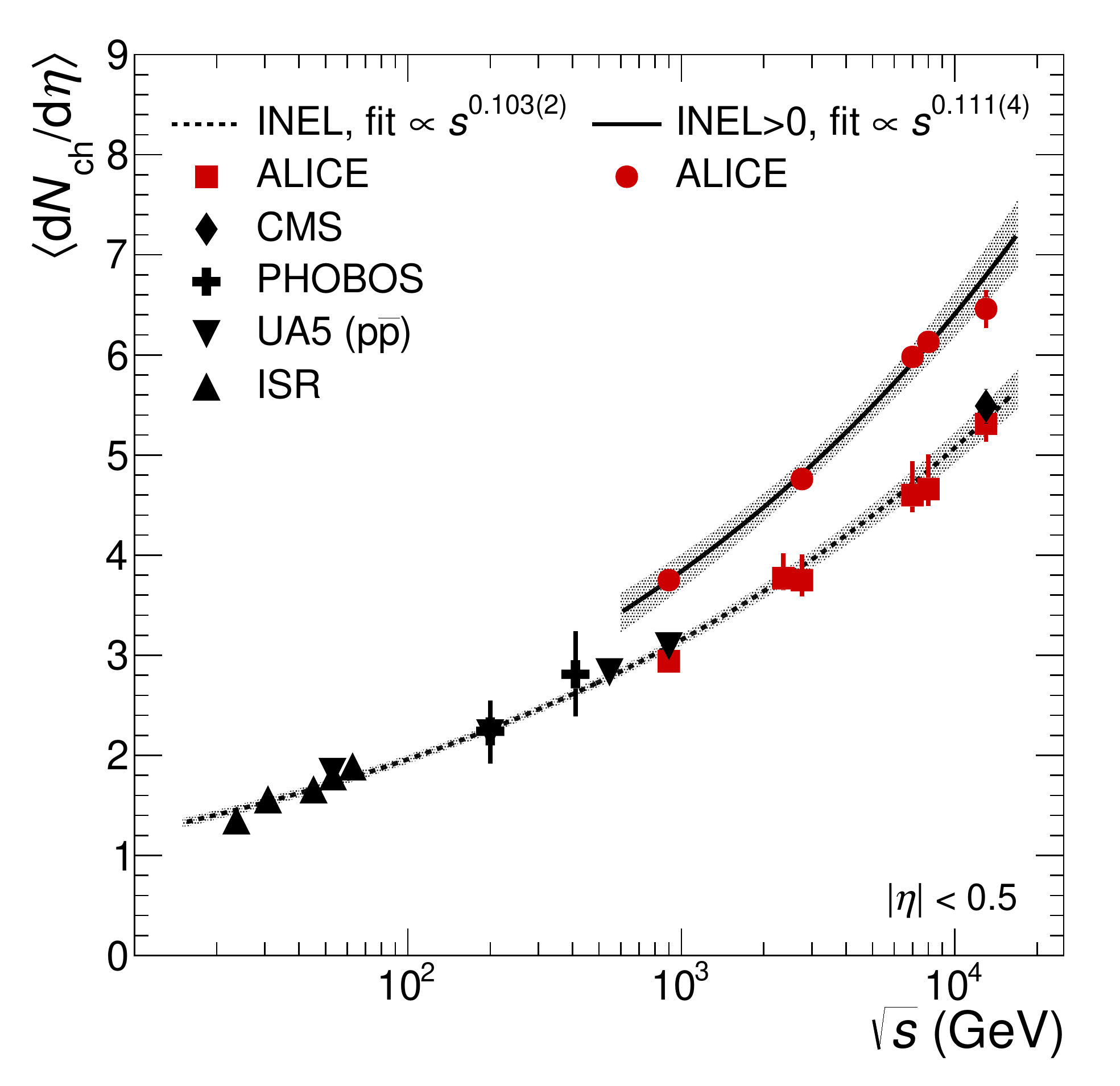}
  \caption{Charged-particle pseudorapidity density measured in the central pseudorapidity region \etaless{0.5} for INEL and \inelgtzero events~\cite{Thome:1977ky,Alner:1986xu,Alpgard:1982zx,Alner:1987wb,Alver:2010ck,Khachatryan:2015jna,Aamodt:2010ft,Aamodt:2010pp,LongMultiPaper}. The uncertainties are the quadratic sum of statistical and systematic contributions. The lines are power-law fits of the energy dependence of the data and the grey bands represent the standard deviation of the fits.}
  \label{fig:sqrts}
\end{figure}

\begin{figure}[t]
  \centering
  \includegraphics[width=0.6\linewidth]{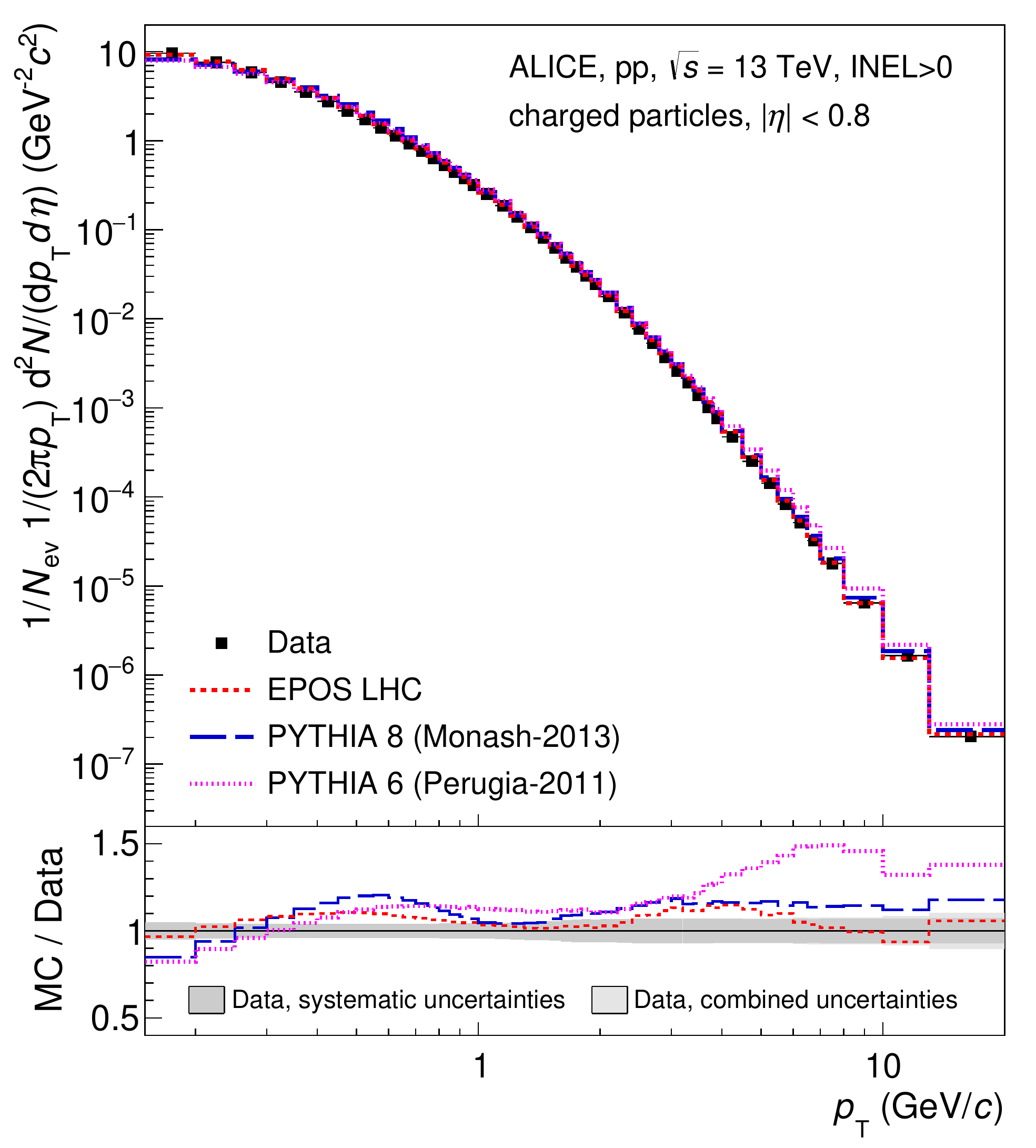}
  \caption{Invariant charged-particle yield as a function of \pt normalised to \inelgtzero events. The data are compared to Monte Carlo calculations~\cite{Sjostrand:2006za,Skands:2010ak,Sjostrand:2014zea,Skands:2014pea,Drescher:2000ha,Pierog:2013ria}. For the ratio of models (MC) and data (lower panel) the systematic and total uncertainties of the data are shown as grey bands.
  }
  \label{fig_pt0} 
\end{figure}

\begin{figure}[t]
  \centering
  \includegraphics[width=0.6\linewidth]{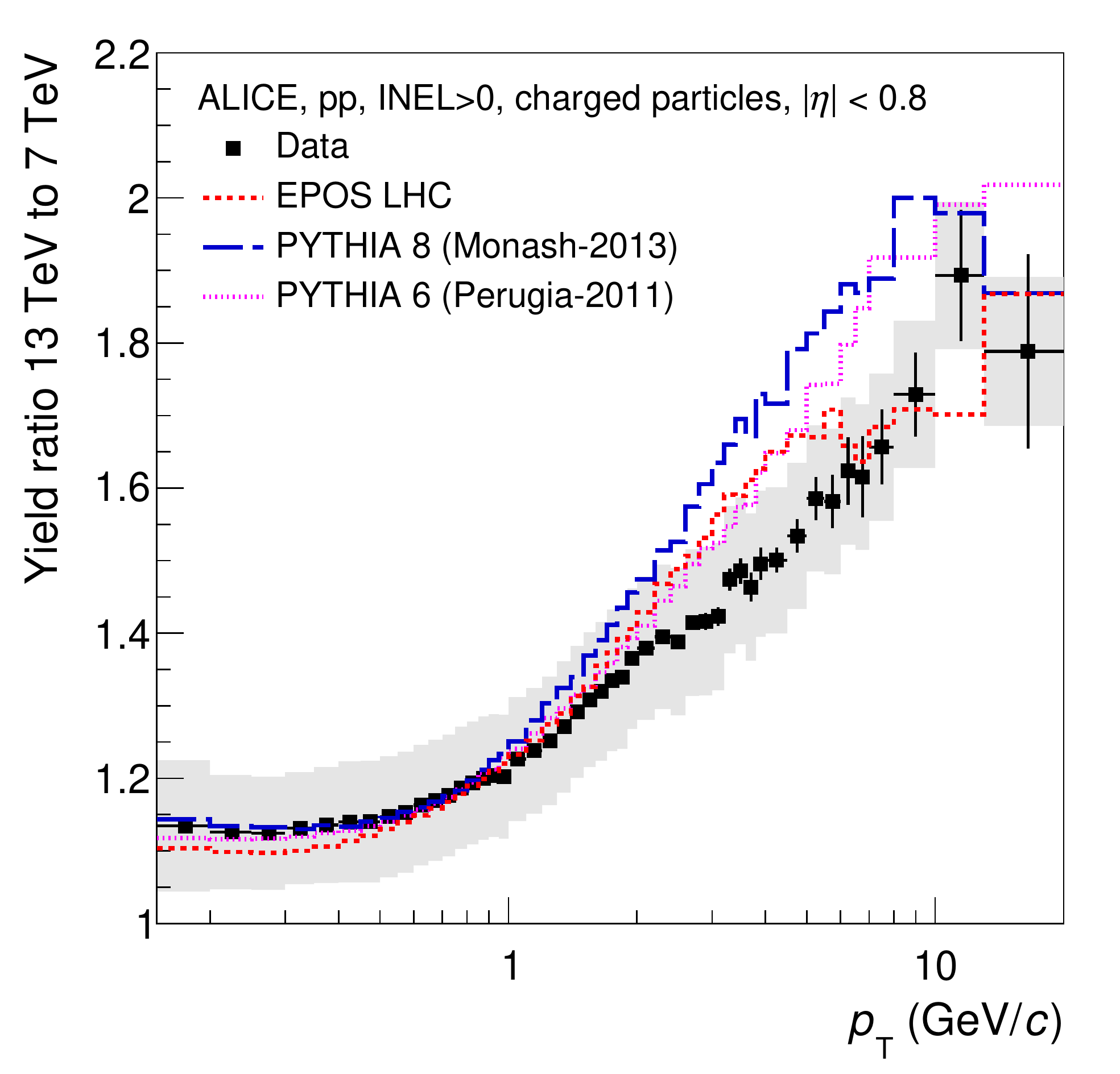}
  \caption{Ratio of transverse-momentum spectra in \inelgtzero events at \sqrts = 13 and 7 TeV. The boxes represent the systematic uncertainties. The data are compared to Monte Carlo calculations~\cite{Sjostrand:2006za,Skands:2010ak,Sjostrand:2014zea,Skands:2014pea,Drescher:2000ha,Pierog:2013ria}. 
  }
  \label{fig_pt1} 
\end{figure}

\begin{figure}[t]
  \centering
  \includegraphics[width=0.6\linewidth]{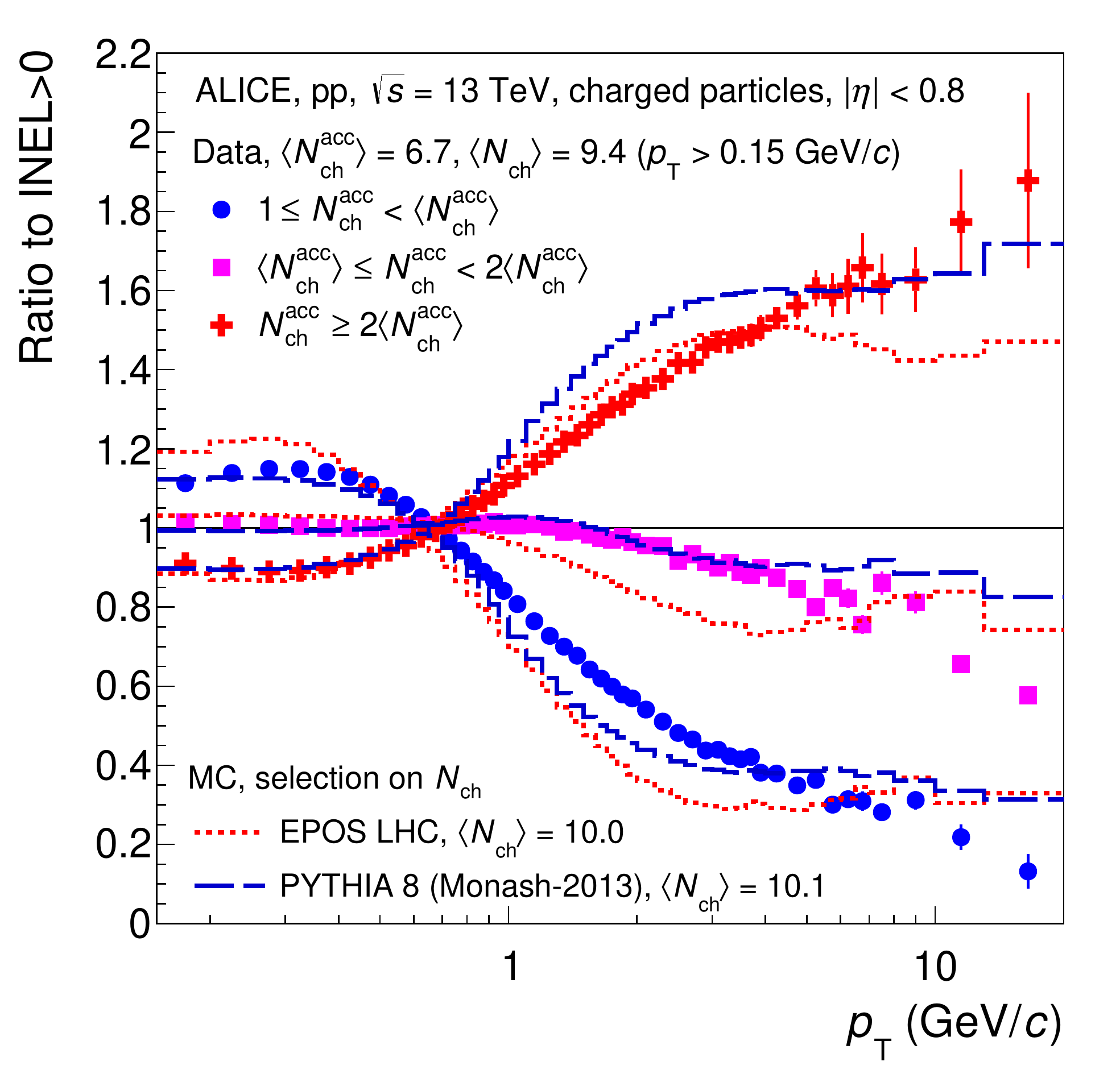}
  \caption{Ratios of transverse-momentum distributions of charged particles 
    in three intervals of multiplicities to the respective one for inclusive (\inelgtzero) collisions.
    The spectra were normalized by the integral prior to division.
    The data are compared to Monte Carlo calculations~\cite{Sjostrand:2014zea,Skands:2014pea,Drescher:2000ha,Pierog:2013ria}. 
  }
  \label{fig_pt3} 
\end{figure}

Figure~\ref{fig:dndeta} shows the average charged-particle density distribution \average{\dndeta} measured in INEL and \inelgtzero events in the pseudorapidity range \etaless{1.8}. The data points have been symmetrised averaging the results obtained in $\pm \eta$, which were consistent within statistical uncertainties. The corresponding pseudorapidity densities in \etaless{0.5} are \DNDETAINEL and \DNDETAINELGTZERO, respectively. The pseudorapidity density for the \inelgtzero events is also measured in \etaless{1} for direct comparison with \inelgtzero results reported by ALICE at lower energies~\cite{Aamodt:2010pp} and is \DNDETAINELGTZEROONE.
Also shown in Fig.~\ref{fig:dndeta} are the results recently published by the CMS Collaboration for inelastic collisions~\cite{Khachatryan:2015jna}, which agree, within the uncertainties, with the measurement presented here. We compared our measurement to Monte Carlo calculations performed with PYTHIA~6~\cite{Sjostrand:2006za} (Perugia-2011~\cite{Skands:2010ak}), PYTHIA~8~\cite{Sjostrand:2014zea} (Monash-2013~\cite{Skands:2014pea}) and EPOS~LHC\footnote{Calculations performed with CRMC package version 1.5.3.}~\cite{Drescher:2000ha,Pierog:2013ria} in both the INEL and \inelgtzero event classes. PYTHIA~6 calculations are in better agreement with the data than PYTHIA~8 in both classes, with PYTHIA~8 being higher than the data by about 12\% (7\%) in INEL events and about 7\% (3\%) in \inelgtzero events at $\eta$ $\sim$ 0 ($\eta \sim 1.5$). EPOS~LHC calculations are about 7\% (4\%) and about 7\% (5\%) higher than the data in INEL and \inelgtzero events, respectively, at $\eta$ $\sim$ 0 ($\eta \sim 1.5$).
In Fig.~\ref{fig:sqrts} we show a compilation of results on pseudorapidity density of charged particles measured in \etaless{0.5} for the INEL and \inelgtzero results at different proton-proton collider energies~\cite{Thome:1977ky,Alner:1986xu,Alpgard:1982zx,Alner:1987wb,Alver:2010ck,Khachatryan:2015jna,Aamodt:2010ft,Aamodt:2010pp,LongMultiPaper}.
The energy dependence of \average{\dndeta} is parametrised by the power law $a s^{b}$ fitted to data, where $a$ and $b$ are free parameters.
By combining the data at lower energies with ALICE and CMS results at \sqrts = 13 TeV, we obtain $b = 0.103 \pm 0.002$ and $b = 0.111 \pm 0.004$ for INEL and \inelgtzero event classes, respectively. Notice that the fit results assume that uncertainties at different centre-of-mass energies are independent, which is not strictly the case.

Figure~\ref{fig_pt0} presents the measured \pt spectrum and its comparison with calculations with PYTHIA~6 (Perugia-2011), PYTHIA~8 (Monash-2013) and EPOS~LHC. For bulk particle production, the mechanism of colour reconnection is an important one in the PYTHIA models (see discussion below and in ref. \cite{Sjostrand:2013cya}). EPOS is a model based on the Gribov-Regge theory at parton level \cite{Drescher:2000ha}. Collective (flow-like) effects are incorporated in the EPOS3 version \cite{Werner:2013tya} and treated via parametrisations in the EPOS~LHC version \cite{Pierog:2013ria}. These event generators, benefitting from the tuning performed on the LHC data in Run 1, describe the \pt spectrum reasonably well, although not in detail. It is interesting to note that both PYTHIA~8 and EPOS~LHC models show a similar pattern in the ratio to data with discrepancies up to 20\% and that PYTHIA~6 overestimates particle production at high \pt.

Figure~\ref{fig_pt1} shows the ratio of transverse-momentum spectra of charged particles at \sqrts = 13 TeV and 7 TeV. The published data at \sqrts = 7 TeV~\cite{Abelev:2013ala} were for INEL events. We have recalculated the normalisation of the spectrum to correspond to \inelgtzero events in a similar manner as done for \sqrts = 13 TeV. The trigger and event-selection efficiency for \inelgtzero events at \sqrts = 7 TeV was estimated using the same Monte Carlo simulations used for the publication~\cite{Abelev:2013ala}.
The systematic uncertainties of the ratio are the quadratic sum of uncertainties at the two energies. As expected, the spectrum is significantly harder at \sqrts = 13 TeV than at \sqrts = 7 TeV. PYTHIA~6, PYTHIA~8 and EPOS~LHC reproduce the trend observed in the data, but exhibit a slightly more pronounced hardening with energy in the transverse momentum region of a few \GeVc. The effect appears to be more significant in PYTHIA~8 than in PYTHIA~6 and EPOS~LHC.

The correlation of the particle mean transverse momentum (\mpt) with the multiplicity of the event (\nch) first observed at the \SPS collider \cite{Arnison:1982ed}, has been studied by many experiments at hadron colliders in pp($\bar{\rm p}$) covering collision energies from \sqrts~=~31~GeV up to 7~TeV~\cite{ABCDHW, Albajar:1989an, PhysRevLett.60.1622,Adams:2006xb,Aaltonen:2009ne,Aamodt:2010my,Khachatryan:2010nk,Aad:2010ac,Abelev:2013bla}. The increase of \mpt with \nch in the central rapidity region observed in all experiments could be reproduced in the PYTHIA event generator only if a mechanism of hadronisation with colour reconnections (CR) is considered \cite{Skands:2007zg,Corke:2010yf,Corke:2011yy,Sjostrand:2013cya}.
A connection between CR and features of collective flow has been conjectured in~\cite{Ortiz:2013yxa}. In heavy-ion collisions, collective flow is established as a genuine space-time evolution of a fireball, while CR in PYTHIA is a mechanism invoked for hadronisation. The relevance of the CR-flow conjecture is currently investigated further~\cite{Bierlich:2015rha}.
A mechanism involving collective string hadronisation is also used in the EPOS model~\cite{Pierog:2013ria}.

Figure~\ref{fig_pt3} shows the ratio of spectra measured in three intervals of multiplicity to the inclusive (\inelgtzero) spectrum. For this ratio, the spectra were normalised by the integral prior to dividing. The selection is performed on the multiplicity measured in the same kinematic region as the spectrum, \etaless{0.8} and 0.15~$<$ \pt $<$~20 \GeVc, using the measured track multiplicity \nchacc for data and the true value of \nch known in Monte Carlo events. For \inelgtzero events, \mnchacc = 6.73 (and, from the spectrum in Fig.~\ref{fig_pt0}, \mnch = 9.41 $\pm$ 0.38) for data and \mnch = 10.13 for PYTHIA~8 and \mnch = 9.97 for EPOS~LHC events. The low-multiplicity interval corresponds to \nch (\nchacc) smaller than the average value in \inelgtzero events, \mnch (\mnchacc), the medium-multiplicity interval covers between \mnch (\mnchacc) and twice \mnch (\mnchacc), while the high-multiplicity interval includes all events with \nch (\nchacc) $\ge$ 2 \mnch (\mnchacc). Given that the measurement efficiency of the \pt spectrum for \inelgtzero events with \nch=1 is about 50\%, the data is slightly biased for the lowest multiplicity interval. This leads to a slight hardening of the measured spectrum, but the magnitude of the spectral shape change, of a few percent, is clearly smaller than the observed difference between data and models. The systematic uncertainties of the measured spectra cancel out completely in the ratios. A residual contribution, not estimated at this stage, is that of the contamination from strange-particle decays.

It is known that the increase of \mpt as a function of multiplicity is moderate \cite{Abelev:2013bla}. The data in Fig.~\ref{fig_pt3} show that the correlation of the spectrum with multiplicity is prominent for the whole \pt range and in particular that it is stronger at high \pt.
In first order, this correlation arises naturally from jets, giving the leading high-\pt hadron and a significant contribution to multiplicity.
The general features seen in the data, which are similar to those first seen at \sqrts =~0.9~TeV \cite{Aamodt:2010my}, are reproduced by PYTHIA~8 and EPOS~LHC fairly well, but some disagreements are noticeable too, in particular in the \pt region of a few \GeVc. This is more prominent for EPOS~LHC. It was shown earlier \cite{Abelev:2013bla} that both EPOS~LHC and PYTHIA~8 reproduce well, although slightly overpredicting, the correlation of \mpt with \nch. The present data on spectral shape highlight some deficiencies in both models concerning the description of spectral shapes as a function of multiplicity.

%% file: conclusions.tex
\section{Conclusions}\label{sec:conclusions}
We have reported the measurement of the pseudorapidity and transverse-momentum distributions of charged particles produced in proton-proton collisions at \sqrts = 13 TeV with the ALICE detector at LHC.
The pseudorapidity distribution is measured for two normalisation classes: inelastic events (INEL) and events having at least one charged particle in the pseudorapidity interval \etaless{1} (\inelgtzero). The charged-particle densities in \etaless{0.5} are \DNDETAINEL and \DNDETAINELGTZERO, respectively.
The transverse-momentum distribution is measured in the range 0.15 $<$ \pt $<$ 20 \GeVc and \etaless{0.8} for \inelgtzero events. The spectrum is significantly harder than at \sqrts = 7 TeV and shows rich features when correlated with the charged-particle multiplicity measured in the same kinematic region.
The results are found to be in fair agreement with the expectations from lower energy extrapolations and with the calculations from PYTHIA and EPOS Monte Carlo generators, but not in all details. Both models exhibit a slightly more pronounced hardening of the \pt distributions with collision energy than the data for transverse momenta above a few \GeVc.

%% file: acknowledgements.tex

The ALICE Collaboration would like to thank all its engineers and technicians for their invaluable contributions to the construction of the experiment and the CERN accelerator teams for the outstanding performance of the LHC complex.
The ALICE Collaboration gratefully acknowledges the resources and support provided by all Grid centres and the Worldwide LHC Computing Grid (WLCG) collaboration.
The ALICE Collaboration acknowledges the following funding agencies for their support in building and
running the ALICE detector:
State Committee of Science,  World Federation of Scientists (WFS)
and Swiss Fonds Kidagan, Armenia;
Conselho Nacional de Desenvolvimento Cient\'{\i}fico e Tecnol\'{o}gico (CNPq), Financiadora de Estudos e Projetos (FINEP),
Funda\c{c}\~{a}o de Amparo \`{a} Pesquisa do Estado de S\~{a}o Paulo (FAPESP);
National Natural Science Foundation of China (NSFC), the Chinese Ministry of Education (CMOE)
and the Ministry of Science and Technology of China (MSTC);
Ministry of Education and Youth of the Czech Republic;
Danish Natural Science Research Council, the Carlsberg Foundation and the Danish National Research Foundation;
The European Research Council under the European Community's Seventh Framework Programme;
Helsinki Institute of Physics and the Academy of Finland;
French CNRS-IN2P3, the `Region Pays de Loire', `Region Alsace', `Region Auvergne' and CEA, France;
German Bundesministerium fur Bildung, Wissenschaft, Forschung und Technologie (BMBF) and the Helmholtz Association;
General Secretariat for Research and Technology, Ministry of Development, Greece;
National Research, Development and Innovation Office (NKFIH), Hungary;
Department of Atomic Energy and Department of Science and Technology of the Government of India;
Istituto Nazionale di Fisica Nucleare (INFN) and Centro Fermi -
Museo Storico della Fisica e Centro Studi e Ricerche ``Enrico Fermi'', Italy;
MEXT Grant-in-Aid for Specially Promoted Research, Ja\-pan;
Joint Institute for Nuclear Research, Dubna;
National Research Foundation of Korea (NRF);
Consejo Nacional de Cienca y Tecnologia (CONACYT), Direccion General de Asuntos del Personal Academico(DGAPA), M\'{e}xico, Amerique Latine Formation academique - 
European Commission~(ALFA-EC) and the EPLANET Program~(European Particle Physics Latin American Network);
Stichting voor Fundamenteel Onderzoek der Materie (FOM) and the Nederlandse Organisatie voor Wetenschappelijk Onderzoek (NWO), Netherlands;
Research Council of Norway (NFR);
National Science Centre, Poland;
Ministry of National Education/Institute for Atomic Physics and National Council of Scientific Research in Higher Education~(CNCSI-UEFISCDI), Romania;
Ministry of Education and Science of Russian Federation, Russian
Academy of Sciences, Russian Federal Agency of Atomic Energy,
Russian Federal Agency for Science and Innovations and The Russian
Foundation for Basic Research;
Ministry of Education of Slovakia;
Department of Science and Technology, South Africa;
Centro de Investigaciones Energeticas, Medioambientales y Tecnologicas (CIEMAT), E-Infrastructure shared between Europe and Latin America (EELA), 
Ministerio de Econom\'{i}a y Competitividad (MINECO) of Spain, Xunta de Galicia (Conseller\'{\i}a de Educaci\'{o}n),
Centro de Aplicaciones Tecnológicas y Desarrollo Nuclear (CEA\-DEN), Cubaenerg\'{\i}a, Cuba, and IAEA (International Atomic Energy Agency);
Swedish Research Council (VR) and Knut $\&$ Alice Wallenberg
Foundation (KAW);
Ukraine Ministry of Education and Science;
United Kingdom Science and Technology Facilities Council (STFC);
The United States Department of Energy, the United States National
Science Foundation, the State of Texas, and the State of Ohio;
Ministry of Science, Education and Sports of Croatia and  Unity through Knowledge Fund, Croatia;
Council of Scientific and Industrial Research (CSIR), New Delhi, India;
Pontificia Universidad Cat\'{o}lica del Per\'{u}.

%% file: Alice_Authorlist_2015-Sep-24.tex


\begingroup
\small
\begin{flushleft}
J.~Adam\Irefn{org40}\And
D.~Adamov\'{a}\Irefn{org84}\And
M.M.~Aggarwal\Irefn{org88}\And
G.~Aglieri Rinella\Irefn{org36}\And
M.~Agnello\Irefn{org110}\And
N.~Agrawal\Irefn{org48}\And
Z.~Ahammed\Irefn{org132}\And
S.U.~Ahn\Irefn{org68}\And
S.~Aiola\Irefn{org136}\And
A.~Akindinov\Irefn{org58}
% ...
\renewcommand\labelenumi{\textsuperscript{\theenumi}~}

\section*{Affiliation notes}
\renewcommand\theenumi{\roman{enumi}}
\begin{Authlist}
\item \Adef{0}Deceased
\item \Adef{idp1753856}{Also at: Georgia State University, Atlanta, Georgia, United States}
\item \Adef{idp3083616}{Also at: Also at Department of Applied Physics, Aligarh Muslim University, Aligarh, India}
\item \Adef{idp3803360}{Also at: M.V. Lomonosov Moscow State University, D.V. Skobeltsyn Institute of Nuclear, Physics, Moscow, Russia}
\end{Authlist}

\section*{Collaboration Institutes}
\renewcommand\theenumi{\arabic{enumi}~}
\begin{Authlist}

\item \Idef{org1}A.I. Alikhanyan National Science Laboratory (Yerevan Physics Institute) Foundation, Yerevan, Armenia
\item \Idef{org2}Benem\'{e}rita Universidad Aut\'{o}noma de Puebla, Puebla, Mexico
\item \Idef{org3}Bogolyubov Institute for Theoretical Physics, Kiev, Ukraine
\item \Idef{org4}Bose Institute, Department of Physics and Centre for Astroparticle Physics and Space Science (CAPSS), Kolkata, India
\item \Idef{org5}Budker Institute for Nuclear Physics, Novosibirsk, Russia
\item \Idef{org6}California Polytechnic State University, San Luis Obispo, California, United States
\item \Idef{org7}Central China Normal University, Wuhan, China
\item \Idef{org8}Centre de Calcul de l'IN2P3, Villeurbanne, France
\item \Idef{org9}Centro de Aplicaciones Tecnol\'{o}gicas y Desarrollo Nuclear (CEADEN), Havana, Cuba
\item \Idef{org10}Centro de Investigaciones Energ\'{e}ticas Medioambientales y Tecnol\'{o}gicas (CIEMAT), Madrid, Spain
\item \Idef{org11}Centro de Investigaci\'{o}n y de Estudios Avanzados (CINVESTAV), Mexico City and M\'{e}rida, Mexico
\item \Idef{org12}Centro Fermi - Museo Storico della Fisica e Centro Studi e Ricerche ``Enrico Fermi'', Rome, Italy
\item \Idef{org13}Chicago State University, Chicago, Illinois, USA
\item \Idef{org14}China Institute of Atomic Energy, Beijing, China
\item \Idef{org15}Commissariat \`{a} l'Energie Atomique, IRFU, Saclay, France
\item \Idef{org16}COMSATS Institute of Information Technology (CIIT), Islamabad, Pakistan
\item \Idef{org17}Departamento de F\'{\i}sica de Part\'{\i}culas and IGFAE, Universidad de Santiago de Compostela, Santiago de Compostela, Spain
\item \Idef{org18}Department of Physics and Technology, University of Bergen, Bergen, Norway
\item \Idef{org19}Department of Physics, Aligarh Muslim University, Aligarh, India
\item \Idef{org20}Department of Physics, Ohio State University, Columbus, Ohio, United States
\item \Idef{org21}Department of Physics, Sejong University, Seoul, South Korea
\item \Idef{org22}Department of Physics, University of Oslo, Oslo, Norway
\item \Idef{org23}Dipartimento di Elettrotecnica ed Elettronica del Politecnico, Bari, Italy
\item \Idef{org24}Dipartimento di Fisica dell'Universit\`{a} 'La Sapienza' and Sezione INFN Rome, Italy
\item \Idef{org25}Dipartimento di Fisica dell'Universit\`{a} and Sezione INFN, Cagliari, Italy
\item \Idef{org26}Dipartimento di Fisica dell'Universit\`{a} and Sezione INFN, Trieste, Italy
\item \Idef{org27}Dipartimento di Fisica dell'Universit\`{a} and Sezione INFN, Turin, Italy
\item \Idef{org28}Dipartimento di Fisica e Astronomia dell'Universit\`{a} and Sezione INFN, Bologna, Italy
\item \Idef{org29}Dipartimento di Fisica e Astronomia dell'Universit\`{a} and Sezione INFN, Catania, Italy
\item \Idef{org30}Dipartimento di Fisica e Astronomia dell'Universit\`{a} and Sezione INFN, Padova, Italy
\item \Idef{org31}Dipartimento di Fisica `E.R.~Caianiello' dell'Universit\`{a} and Gruppo Collegato INFN, Salerno, Italy
\item \Idef{org32}Dipartimento di Scienze e Innovazione Tecnologica dell'Universit\`{a} del  Piemonte Orientale and Gruppo Collegato INFN, Alessandria, Italy
\item \Idef{org33}Dipartimento Interateneo di Fisica `M.~Merlin' and Sezione INFN, Bari, Italy
\item \Idef{org34}Division of Experimental High Energy Physics, University of Lund, Lund, Sweden
\item \Idef{org35}Eberhard Karls Universit\"{a}t T\"{u}bingen, T\"{u}bingen, Germany
\item \Idef{org36}European Organization for Nuclear Research (CERN), Geneva, Switzerland
\item \Idef{org37}Excellence Cluster Universe, Technische Universit\"{a}t M\"{u}nchen, Munich, Germany
\item \Idef{org38}Faculty of Engineering, Bergen University College, Bergen, Norway
\item \Idef{org39}Faculty of Mathematics, Physics and Informatics, Comenius University, Bratislava, Slovakia
\item \Idef{org40}Faculty of Nuclear Sciences and Physical Engineering, Czech Technical University in Prague, Prague, Czech Republic
\item \Idef{org41}Faculty of Science, P.J.~\v{S}af\'{a}rik University, Ko\v{s}ice, Slovakia
\item \Idef{org42}Faculty of Technology, Buskerud and Vestfold University College, Vestfold, Norway
\item \Idef{org43}Frankfurt Institute for Advanced Studies, Johann Wolfgang Goethe-Universit\"{a}t Frankfurt, Frankfurt, Germany
\item \Idef{org44}Gangneung-Wonju National University, Gangneung, South Korea
\item \Idef{org45}Gauhati University, Department of Physics, Guwahati, India
\item \Idef{org46}Helsinki Institute of Physics (HIP), Helsinki, Finland
\item \Idef{org47}Hiroshima University, Hiroshima, Japan
\item \Idef{org48}Indian Institute of Technology Bombay (IIT), Mumbai, India
\item \Idef{org49}Indian Institute of Technology Indore, Indore (IITI), India
\item \Idef{org50}Inha University, Incheon, South Korea
\item \Idef{org51}Institut de Physique Nucl\'eaire d'Orsay (IPNO), Universit\'e Paris-Sud, CNRS-IN2P3, Orsay, France
\item \Idef{org52}Institut f\"{u}r Informatik, Johann Wolfgang Goethe-Universit\"{a}t Frankfurt, Frankfurt, Germany
\item \Idef{org53}Institut f\"{u}r Kernphysik, Johann Wolfgang Goethe-Universit\"{a}t Frankfurt, Frankfurt, Germany
\item \Idef{org54}Institut f\"{u}r Kernphysik, Westf\"{a}lische Wilhelms-Universit\"{a}t M\"{u}nster, M\"{u}nster, Germany
\item \Idef{org55}Institut Pluridisciplinaire Hubert Curien (IPHC), Universit\'{e} de Strasbourg, CNRS-IN2P3, Strasbourg, France
\item \Idef{org56}Institute for Nuclear Research, Academy of Sciences, Moscow, Russia
\item \Idef{org57}Institute for Subatomic Physics of Utrecht University, Utrecht, Netherlands
\item \Idef{org58}Institute for Theoretical and Experimental Physics, Moscow, Russia
\item \Idef{org59}Institute of Experimental Physics, Slovak Academy of Sciences, Ko\v{s}ice, Slovakia
\item \Idef{org60}Institute of Physics, Academy of Sciences of the Czech Republic, Prague, Czech Republic
\item \Idef{org61}Institute of Physics, Bhubaneswar, India
\item \Idef{org62}Institute of Space Science (ISS), Bucharest, Romania
\item \Idef{org63}Instituto de Ciencias Nucleares, Universidad Nacional Aut\'{o}noma de M\'{e}xico, Mexico City, Mexico
\item \Idef{org64}Instituto de F\'{\i}sica, Universidad Nacional Aut\'{o}noma de M\'{e}xico, Mexico City, Mexico
\item \Idef{org65}iThemba LABS, National Research Foundation, Somerset West, South Africa
\item \Idef{org66}Joint Institute for Nuclear Research (JINR), Dubna, Russia
\item \Idef{org67}Konkuk University, Seoul, South Korea
\item \Idef{org68}Korea Institute of Science and Technology Information, Daejeon, South Korea
\item \Idef{org69}KTO Karatay University, Konya, Turkey
\item \Idef{org70}Laboratoire de Physique Corpusculaire (LPC), Clermont Universit\'{e}, Universit\'{e} Blaise Pascal, CNRS--IN2P3, Clermont-Ferrand, France
\item \Idef{org71}Laboratoire de Physique Subatomique et de Cosmologie, Universit\'{e} Grenoble-Alpes, CNRS-IN2P3, Grenoble, France
\item \Idef{org72}Laboratori Nazionali di Frascati, INFN, Frascati, Italy
\item \Idef{org73}Laboratori Nazionali di Legnaro, INFN, Legnaro, Italy
\item \Idef{org74}Lawrence Berkeley National Laboratory, Berkeley, California, United States
\item \Idef{org75}Moscow Engineering Physics Institute, Moscow, Russia
\item \Idef{org76}Nagasaki Institute of Applied Science, Nagasaki, Japan
\item \Idef{org77}National Centre for Nuclear Studies, Warsaw, Poland
\item \Idef{org78}National Institute for Physics and Nuclear Engineering, Bucharest, Romania
\item \Idef{org79}National Institute of Science Education and Research, Bhubaneswar, India
\item \Idef{org80}National Research Centre Kurchatov Institute, Moscow, Russia
\item \Idef{org81}Niels Bohr Institute, University of Copenhagen, Copenhagen, Denmark
\item \Idef{org82}Nikhef, Nationaal instituut voor subatomaire fysica, Amsterdam, Netherlands
\item \Idef{org83}Nuclear Physics Group, STFC Daresbury Laboratory, Daresbury, United Kingdom
\item \Idef{org84}Nuclear Physics Institute, Academy of Sciences of the Czech Republic, \v{R}e\v{z} u Prahy, Czech Republic
\item \Idef{org85}Oak Ridge National Laboratory, Oak Ridge, Tennessee, United States
\item \Idef{org86}Petersburg Nuclear Physics Institute, Gatchina, Russia
\item \Idef{org87}Physics Department, Creighton University, Omaha, Nebraska, United States
\item \Idef{org88}Physics Department, Panjab University, Chandigarh, India
\item \Idef{org89}Physics Department, University of Athens, Athens, Greece
\item \Idef{org90}Physics Department, University of Cape Town, Cape Town, South Africa
\item \Idef{org91}Physics Department, University of Jammu, Jammu, India
\item \Idef{org92}Physics Department, University of Rajasthan, Jaipur, India
\item \Idef{org93}Physik Department, Technische Universit\"{a}t M\"{u}nchen, Munich, Germany
\item \Idef{org94}Physikalisches Institut, Ruprecht-Karls-Universit\"{a}t Heidelberg, Heidelberg, Germany
\item \Idef{org95}Purdue University, West Lafayette, Indiana, United States
\item \Idef{org96}Pusan National University, Pusan, South Korea
\item \Idef{org97}Research Division and ExtreMe Matter Institute EMMI, GSI Helmholtzzentrum f\"ur Schwerionenforschung, Darmstadt, Germany
\item \Idef{org98}Rudjer Bo\v{s}kovi\'{c} Institute, Zagreb, Croatia
\item \Idef{org99}Russian Federal Nuclear Center (VNIIEF), Sarov, Russia
\item \Idef{org100}Saha Institute of Nuclear Physics, Kolkata, India
\item \Idef{org101}School of Physics and Astronomy, University of Birmingham, Birmingham, United Kingdom
\item \Idef{org102}Secci\'{o}n F\'{\i}sica, Departamento de Ciencias, Pontificia Universidad Cat\'{o}lica del Per\'{u}, Lima, Peru
\item \Idef{org103}Sezione INFN, Bari, Italy
\item \Idef{org104}Sezione INFN, Bologna, Italy
\item \Idef{org105}Sezione INFN, Cagliari, Italy
\item \Idef{org106}Sezione INFN, Catania, Italy
\item \Idef{org107}Sezione INFN, Padova, Italy
\item \Idef{org108}Sezione INFN, Rome, Italy
\item \Idef{org109}Sezione INFN, Trieste, Italy
\item \Idef{org110}Sezione INFN, Turin, Italy
\item \Idef{org111}SSC IHEP of NRC Kurchatov institute, Protvino, Russia
\item \Idef{org112}Stefan Meyer Institut f\"{u}r Subatomare Physik (SMI), Vienna, Austria
\item \Idef{org113}SUBATECH, Ecole des Mines de Nantes, Universit\'{e} de Nantes, CNRS-IN2P3, Nantes, France
\item \Idef{org114}Suranaree University of Technology, Nakhon Ratchasima, Thailand
\item \Idef{org115}Technical University of Ko\v{s}ice, Ko\v{s}ice, Slovakia
\item \Idef{org116}Technical University of Split FESB, Split, Croatia
\item \Idef{org117}The Henryk Niewodniczanski Institute of Nuclear Physics, Polish Academy of Sciences, Cracow, Poland
\item \Idef{org118}The University of Texas at Austin, Physics Department, Austin, Texas, USA
\item \Idef{org119}Universidad Aut\'{o}noma de Sinaloa, Culiac\'{a}n, Mexico
\item \Idef{org120}Universidade de S\~{a}o Paulo (USP), S\~{a}o Paulo, Brazil
\item \Idef{org121}Universidade Estadual de Campinas (UNICAMP), Campinas, Brazil
\item \Idef{org122}University of Houston, Houston, Texas, United States
\item \Idef{org123}University of Jyv\"{a}skyl\"{a}, Jyv\"{a}skyl\"{a}, Finland
\item \Idef{org124}University of Liverpool, Liverpool, United Kingdom
\item \Idef{org125}University of Tennessee, Knoxville, Tennessee, United States
\item \Idef{org126}University of the Witwatersrand, Johannesburg, South Africa
\item \Idef{org127}University of Tokyo, Tokyo, Japan
\item \Idef{org128}University of Tsukuba, Tsukuba, Japan
\item \Idef{org129}University of Zagreb, Zagreb, Croatia
\item \Idef{org130}Universit\'{e} de Lyon, Universit\'{e} Lyon 1, CNRS/IN2P3, IPN-Lyon, Villeurbanne, France
\item \Idef{org131}V.~Fock Institute for Physics, St. Petersburg State University, St. Petersburg, Russia
\item \Idef{org132}Variable Energy Cyclotron Centre, Kolkata, India
\item \Idef{org133}Warsaw University of Technology, Warsaw, Poland
\item \Idef{org134}Wayne State University, Detroit, Michigan, United States
\item \Idef{org135}Wigner Research Centre for Physics, Hungarian Academy of Sciences, Budapest, Hungary
\item \Idef{org136}Yale University, New Haven, Connecticut, United States
\item \Idef{org137}Yonsei University, Seoul, South Korea
\item \Idef{org138}Zentrum f\"{u}r Technologietransfer und Telekommunikation (ZTT), Fachhochschule Worms, Worms, Germany
\end{Authlist}
\endgroup